\documentclass[pra,twocolumn,amsmath,amssymb]{revtex4}
\usepackage{amsmath}
\usepackage{graphicx}
\usepackage{rotating}
\usepackage{units}
\usepackage{color}
\usepackage{ulem}
\usepackage{float}
\usepackage[utf8]{inputenc}

\DeclareMathOperator{\sign}{sign}

\graphicspath{ {./Figures/} }

\begin{document}

\title{Quantum optimal control within the rotating wave approximation}

\author{Maximilian Keck, Matthias M. M\"uller, Tommaso Calarco, and Simone Montangero}
\affiliation{Center for Integrated Quantum Science and Technology, Institute for Complex Quantum Systems, University of Ulm, Albert-Einstein-Allee 11, D-89069 Ulm, Germany}

\begin{abstract}
We study the interplay between rotating wave approximation and optimal control. In particular, we show that for a wide class of optimal control problems one can choose the control field such that the Hamiltonian becomes time-independent under the rotating wave approximation. Thus, we show how to recast the functional minimization defined by the optimal control problem into a simpler multi-variable function minimization. We provide the analytic solution to the state-to-state transfer of the paradigmatic two-level system and to the more general star configuration of an $N$-level system. We demonstrate numerically the usefulness of this approach in the more general 
class of connected acyclic $N$-level systems with random spectra. Finally, we use it to design a protocol to entangle Rydberg via constant laser pulses atoms in an experimentally relevant range of parameters. 
\end{abstract}
\maketitle
The Rotating Wave Approximation (RWA) plays a major role in simplifying the quantum mechanical description of laser driven systems in the low intensity regime: it takes into account only the co-rotating field 
with the system and it neglects the counter-rotating part~\cite{Schleich2011,Whaley1984}.  This approximation has been introduced for two level quantum systems, and then generalized for $N$-level systems~\cite{Einwohner1976}. The deviations from this approximation for big intensities are well known and commonly described as Bloch-Siegert shifts, breaking the harmonicity of the system dynamics~\cite{Bloch1940}. Finally, a more general description is given by Floquet theory that allows to treat periodically driven systems~\cite{Verdeny}.

Developing error-free protocols for the manipulation of quantum systems -- also along the development of quantum technologies but not restricted to them -- is one of the major challenges in contemporary research in atom and molecular physics~\cite{Atomsandmolecules}. During the last decades, an increasing contribution in such effort has come from the exploitation of Quantum Optimal Control (QOC), the search for an optimal control pulse to perform a given system manipulation~\cite{Brif2010}. 
Methods to solve QOC problems have been developed~\cite{CRAB, GRAPE, Krotov} and experiments have shown the great benefit from them, see e.g.~\cite{Waldherr2014, Frank2014, Dolde2013, Scheuer2014, Lovecchio2014}.  We have now deep theoretical understanding of QOC, in particular about the possibilities and hurdles to control quantum systems~\cite{Schirmer2001, Turinici2003, D'Alessandro2007},
and we even start to understand the complexity of QOC problems~\cite{Lloyd2014,Caneva2014}.
Graph theory concepts have been exploited to attack a question that lies at the heart of controllability studies: given a Hamiltonian depending on some time dependent tunable control field, is it possible to dynamically connect every pair of arbitrary initial and final state, i.e.\ it is possible to realize every possible state-to-state transfer? A widely used criterion to answer to this question is via dynamical Lie-Algebras~\cite{Schirmer2001}, while here we make use of a different criterion based on graph theory: Turinici and Rabitz showed that if the graph corresponding to the control Hamiltonian is connected and the spacings of the eigenvalues of the uncontrolled part of the Hamiltonian are non-degenerate, then wave function controllability is guaranteed~\cite{Turinici2003}.

In this paper, we identify a significant class of QOC problems where the RWA is applicable and show that within this class QOC problems can be easily solved: the functional optimization can be recast into a multi-variable function minimization, thus simplifying the numerical efforts and improving the theoretical understanding of the process. For this purpose we employ concepts from graph theory to analyze the system dynamics in an easier and more practical picture~\cite{GraphTheory}. This allows us to straightforwardly identify the cases where it is possible to map the time-dependent Hamiltonian onto a time-independent one and to find the solution of the QOC problem. 
As a result of this analysis, we can show that in this class of QOC problems the lower bound for the number of independent control parameters necessary for a successful control of the system -- introduced in a recent work -- is saturated~\cite{Lloyd2014}. 

Finally, as an example of possible applications of this approach we use it to design a protocol to entangle Rydberg atoms~\cite{SaffmanReview}. Indeed, in the last years experiments with Rydberg atoms have attracted increasing attention as a promising platform for implementing quantum information processing algorithms, such as CNOT-gates~\cite{Isenhower,Zhang2012}, two-body~\cite{Gaetan,Urban,Zhang2010} or many-body entanglement~\cite{Saffman2009}.

The paper is structured as follow:  In Section~\ref{sec:Model} we present the model studied and for the sake of completeness we review the work of Einwohner \textit{et al.} on the generalized $N$-level RWA~\cite{Einwohner1976} and
in Section~\ref{sec:graphs} we review the algorithm based on graph theory to recast systems in a time-independent form. Section~\ref{sec:Analytics} presents the analytic solutions of the state-to-state transfer in the two-level system and a special case of the $N$-level system while numerical results are presented in Section~\ref{sec:Numerics}. In Section~\ref{sec:Example} we demonstrate the methods presented here by designing a protocol entangling Rydberg atoms. Finally, in Section~\ref{sec:DiscussionConclusion} we discuss the results and present the conclusions of this work.
\section{The Model}
\label{sec:Model}
In the standard RWA setting one considers a system described by a $N$-dimensional Hamiltonian ($\hbar \equiv 1 $)
\begin{equation}
\label{eq:system}
\hat{H} = \hat{H}_D + \sum\limits_{f=1}^{F} \operatorname{Re} \left( A_f\, e^{-i\omega_f t}\right) \hat{H}_C,
\end{equation}
where $\hat{H}_D$ is the time-independent drift Hamiltonian with eigenvectors $|n\rangle$ and corresponding eigenvalues $E_n\ (n = 0, 1, \dots N-1)$ and $\hat{H}_C$ is the control Hamiltonian, which might correspond to different physical scenarios. Here, we only assume that the diagonal elements of $\hat{H}_{C}$ are all zero. The  coefficients $A_f \in \mathbb{C}$ represent the independent control parameters and $\omega_f \in \mathbb{R}$ the driving frequencies. A relevant example is naturally encountered in molecular or atomic physics when describing the interaction between matter and light in the dipole approximation, where each control with frequency $\omega_f$ and strength $A_f$ is typically realized by a laser and $\hat{H}_C$ is the dipole operator.
In the following we assume that no two driving frequencies are in approximate resonance to the same transition frequency $E_{kj} \equiv E_k - E_j$ and on the other hand every driving frequency is in approximate resonance to at least one transition frequency. There is not much loss of generality in this when considering a QOC problem since this only means that we do not have off resonant controls nor two controls affecting the same transition.
Assuming that the transition frequencies are not degenerate ($E_{kj} \neq E_{k'j'}\ \forall\ (k, j) \neq (k',j')$) we can make use of results presented in Ref.~\cite{Turinici2003} that ensure wave function controllability.
The last assumption we make is that the RWA is valid, i.e. we are in a setting of low intensities and resonant driving frequencies. That is we assume for each set of transition frequency $E_{kj}$, with driving frequency $\omega_f$ in approximate resonance to that transition, and the corresponding amplitude $A_f$ the inequalities
\begin{align}
\begin{split}
\omega_f &\gg \Delta_{kj} \equiv |E_{kj} - \omega_f|\\
\omega_f &\gg A_f,
\end{split}
\end{align}
to hold.\\
Expanding the wave function $\left| \psi (t)\right\rangle$ as
\begin{equation}
\label{eq:expansion}
\left| \psi (t) \right\rangle = \sum\limits_{k=0}^{N-1} c_{k}(t) e^{-i E_k t} \left| k \right\rangle,
\end{equation}
the Schr\"odinger equation
\begin{equation}
i \frac{d}{d t} \left| \psi (t) \right\rangle = \hat{H} \left| \psi (t) \right\rangle
\end{equation}
can be rewritten as a differential equation for the coefficients $\vec{c} = \left(c_0, c_1, \dots , c_{N-1} \right)$:
\begin{equation}
\label{eq:diffeq}
i \dot{c}_k(t) = \sum\limits_{j=0}^{N-1} H^{\text{(I)}}_{kj} c_k(t),
\end{equation}
where 
\begin{equation}
H^{\text{(I)}}_{kj} = \frac{1}{2} \sum\limits_{f=1}^{F} A_f e^{i (E_{kj} - \omega_f) t} + A_f^{*} e^{i (E_{kj} + \omega_f) t} (H_C)_{kj}.
\label{eq:RWA4}
\end{equation}
Following \cite{Einwohner1976}, the multilevel rotating wave approximation is done by neglecting all non-resonant terms $\hat{H}^{\text{(I)}}_{kj}$. This includes far more than just dropping the counter-rotating term (one of the terms inside the sum of Eq.~\eqref{eq:RWA4}), as in the more common RWA for two-level systems, but also all terms of the sum of frequencies that are non-resonant to the transition frequency corresponding to that matrix element, reducing the whole sum to only one element.\\
The resulting matrix elements are denoted by $H^{\text{(II)}}_{kj} \equiv  M^{\text{(II)}}_{kj} e^{it \Delta_{kj}}$, with
\begin{equation}
M^{\text{(II)}}_{kj} = \frac{1}{4} \left[(1+\sign{(E_{kj}})) A_{f} + (1-\sign{(E_{kj}})) A_{f}^{*}\right].
\end{equation}
A change of basis $b_k(t) = e^{i \gamma_k t} c_k(t),\, \gamma_k \in \mathbb{R}$ which has no a priori physical meaning but is merely a mathematical tool transforms Eq.~\eqref{eq:diffeq} into
\begin{equation}
i \dot{b}_k(t) = \sum\limits_{j=0}^{N-1} \left( M^{\text{(II)}}_{kj} - \gamma_k \delta_{kj} \right) e^{i (\gamma_k - \gamma_j + \Delta_{kj}) t} b_k(t).
\end{equation}
To end up with a time-independent operator one has to set all phases equal to zero, that is all $\frac{N}{2}(N-1)$ equations
\begin{equation}
\label{eq:lineqs}
(\gamma_k - \gamma_j + \Delta_{kj}) (H_c)_{kj} = 0,\ (k \neq j)
\end{equation}
have to be solved by choosing an appropriate $\vec{\gamma} = (\gamma_1, \gamma_2, \dots , \gamma_k)$. The resulting system has the solution:
\begin{equation}
\vec{b} (t) = \exp \left\{ -i \left(M^{\text{(II)}} - \text{diag}(\vec{\gamma})\right) t \right\} \vec{b} (t=0).
\end{equation}
The term $(H_c)_{kj}$ is included in \eqref{eq:lineqs} since only for non-vanishing matrix elements the corresponding equation has to be considered.
In general those equations \eqref{eq:lineqs} cannot be solved consistently, but the zeros in $(H_c)_{kj}$ decrease the number of equations to be solved. This transformation has the advantage of possibly making the description of the system simpler. In particular for acyclic graphs we will see that system can always be reduced to a time-independent one.
\section{Time-independent description for acyclic graphs}
\label{sec:graphs}
\begin{figure}[H]
\begin{center}
\setlength{\unitlength}{.9cm}
{\includegraphics[scale=0.3]{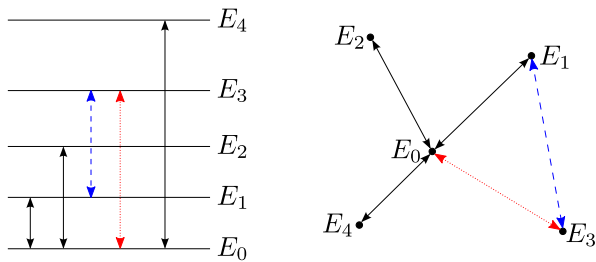}}
\caption{(Color online) Energy level scheme of an example five-level system (left-hand side) and the corresponding graph (right-hand side). The graph is connected but has a cycle ($E_0,E_1,E_3$). Removing the red dotted edge between the vertices $E_0$ and $E_3$ makes the graph acyclic while maintaining the connected property. Removing the blue dashed edge between $E_1$ and $E_3$ also leaves the graph connected and acyclic, but in addition that graph is a star, a type of graph which is studied in Sec.~\ref{sec:N-level}.
}
\label{fig:level_scheme_and_graph}
\end{center}
\end{figure}

Every Hamiltonian consisting of a diagonal drift term $\hat{H}_D$ and a control term $\hat{H}_C$ has a pictorial representation via its energy level scheme, which can easily be mapped onto an undirected graph, see Fig.~\ref{fig:level_scheme_and_graph} for a descriptive example. Every state $\{E_j\}$ is portrayed by a vertex and edges of the graph illustrate non-zero transition elements.
There is a specific class of graphs (and therefore control Hamiltonians), that deserve a special treatment due to their convenient properties. Those are the subset of connected, acyclic graphs (``trees''), where from every vertex $k$ to every other vertex $j$ there exists a unique path, so no cycles are present. In Fig.~\ref{fig:level_scheme_and_graph} removing the blue dashed edge between $E_1$ and $E_3$ or the red dotted edge between $E_0$ and $E_3$ from the complete graph results in an acyclic graph.
As already mentioned in the introduction, connected, acyclic graphs form a very significant subset of QOC Hamiltonians: They connect the complete system dynamics with the smallest number of control parameters: all complex amplitudes $A_f$ and frequencies $\omega_f$ are independent parameters to control.
As a consequence the number of controls is set to $F=N-1$ and the number of equations to solve reduces to $N-1$, so a solution for the equations \eqref{eq:lineqs} always exist.
In the following we briefly explain a constructive approach to find a solution, referring again to Ref.~\cite{Einwohner1976}:
\begin{enumerate}
\item Every pendant vertex has a unique ``successor'' vertex, connected to it by an edge
\item Delete recursively all pendant vertices until one ends up with one vertex $l$, assign an arbitrary real value $\gamma_{l}$ to it.
\item Every vertex $k$ is assigned a value $\gamma_k = \gamma_j - \Delta_{kj}$, where $j$ is the successor of $k$.
\end{enumerate}
The algorithm will be implicitly used in Sec.~\ref{sec:2-level} and \ref{sec:N-level}  and is directly implemented in the numerical calculations presented in Sec.~\ref{sec:N-level}.

\begin{figure*}[htp]
\begin{center}
\setlength{\unitlength}{.9cm}
{\includegraphics[scale=0.19]{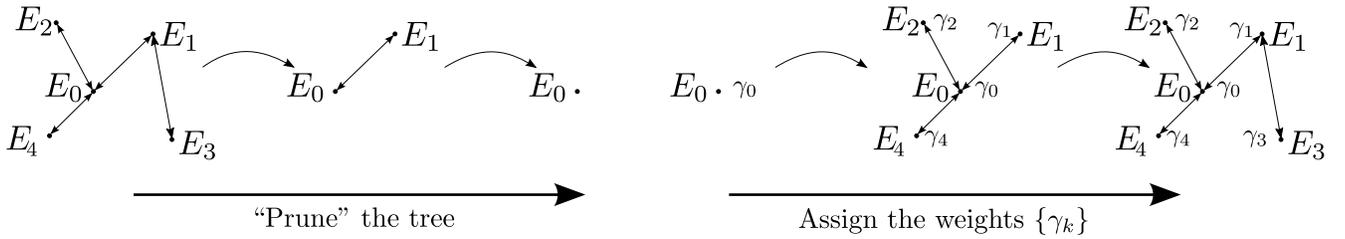}}
\caption{Graph corresponding to an energy scheme of a five-level system taken from Fig.~\ref{fig:level_scheme_and_graph} except for the red dotted edge. The first three figures show the ``pruning'' of the tree, where all pendant vertices and the edges connecting to them are deleted recursively until one root ($E_0$) is left. The last three figures depict how the weights $\{\gamma_k\}$ are recursively assigned to the edges: First the root $E_0$ is assigned an arbitrary value $\gamma_0$, then the three neighboring vertices $E_1, E_2, E_4$ are assigned their values $\gamma_k = \gamma_0 - \Delta_{k0},\, k=1,2,4$ and at last one assigns $E_3$ its value $\gamma_3 = \gamma_1 - \Delta_{31}$.
}
\label{fig:algorithm}
\end{center}
\end{figure*}

To elucidate this algorithm we depicted an example based on the graph from Fig.~\ref{fig:level_scheme_and_graph} without the red dotted edge (see Fig.~\ref{fig:algorithm}). The first three pictures illustrate the ``pruning'' of the tree, that is finding the root of the tree. That is done by finding all pendant vertices of the graph, i.e.\ those that are only connected via one edge, which are in this case the vertices $E_2, E_3$ and $E_4$, and removing them. Repeating this step, that is removing $E_1$, leaves us with the root graph: $E_0$. In the last three pictures we show how the variables $\gamma_k$ are chosen. This is done by assigning an arbitrary real weight $\gamma_0$ to $E_0$ and then re-adding the pendant vertices $E_1, E_2$ and $E_4$ with their corresponding weights $\gamma_k = \gamma - \Delta_{k0}, k = 1, 2, 4$. The algorithm finishes with adding the last vertex $E_3$ with its weight $\gamma_3 = \gamma_1 - \Delta_{31}$.
\section{Analytical results}
\label{sec:Analytics}
In the following we present two systems that have a simple analytic solution: the paradigmatic two-level system and a $N$-level system whose graph is a star. Those solutions provide a general insight in the structure of dynamics and hence the possibilities for QOC.
The algorithm reviewed in Sec.~\ref{sec:graphs} is implicitly used in the following parts.
\subsection{State-to-state transfer for the two-level system}
\label{sec:2-level}
As a starting point of our analysis, we specialize our investigation to the case where $N=2$ and consequently $F=1$. In this case the system given by \eqref{eq:system} takes the form
\begin{align}
\hat{H} &= \hat{H}_D + \operatorname{Re} \left( A\, e^{i \omega t} \right) \hat{H}_C,\\
\label{eq:2_level_system}
\hat{H}_C &= \left|0\right\rangle \left\langle 1 \right| + \left|1\right\rangle \left\langle 0 \right|,
\end{align}
where the indices of $A \equiv A_1$, $\omega \equiv \omega_1$ and $\Delta \equiv \Delta_{01}$ were dropped for clarity. Employing the RWA and assuming the system initially to be in the ground state $\vec{c}\,(t=0) = (1,0)$, one recovers the known solution
\begin{align}
c_0(t)& = e^{i\Delta /2 t}\left( \cos \frac{\tilde{A}}{2} t - i \frac{\Delta}{\tilde{A}} \sin  \frac{\tilde{A}}{2} t \right)\\
c_1(t) &= e^{-i\Delta /2 t} \frac{A}{\tilde{A}} \sin \frac{\tilde{A}}{2} t,
\end{align}
where $\tilde{A} \equiv \sqrt{\Delta^2 + |A|^2}$.
Using the Bloch sphere representation and $\vec{c}_{\text{G}} (\Theta, \phi) = \left( \cos \frac{\Theta}{2}, \sin \frac{\Theta}{2} e^{i \phi} \right), \phi \in [0,2\pi], \Theta \in [0,\pi]$ as the goal state for a state-to-state transfer, we find the equations
\begin{align}
e^{i\Delta /2 t}\left( \cos \frac{\tilde{A}}{2} t - i \frac{\Delta}{\tilde{A}} \sin  \frac{\tilde{A}}{2} t \right) &=  \cos \frac{\Theta}{2}\\
e^{-i\Delta /2 t} \frac{A}{\tilde{A}} \sin \frac{\tilde{A}}{2} t &= \sin \frac{\Theta}{2} e^{i\phi}.
\label{eq:2level}
\end{align}
From Eq.~\eqref{eq:2level} one can find that for a given goal state the tunable parameters of the system have to satisfy the inequality
\begin{equation}
\label{eq:2_reachable}
2 \arcsin \frac{|A|}{\tilde{A}} \geq \Theta
\end{equation}
and that the total time $T$ is 
\begin{equation}
\label{eq:2_time}
T = \frac{2}{\tilde{A}} \arcsin \left( \frac{\tilde{A}}{|A|} \sin \frac{\Theta}{2} \right).
\end{equation}
This time $T$ is known in the literature as the quantum speed limit, that is the smallest time necessary to evolve in Hilbert space from the initial state to the goal state at a given fixed energy~\cite{QSL}.
Using Eq.~\eqref{eq:2level} and comparing the complex phases we see that the phase $\alpha$ of $A \equiv |A| e^{i \alpha}$ has to be chosen as
\begin{equation}
\label{eq:2_phase}
\alpha = \phi + \frac{1}{2} \Delta\, T.
\end{equation}
In summary the three Eqs.~\eqref{eq:2_reachable}, \eqref{eq:2_time} and \eqref{eq:2_phase} yield all information necessary to control the system:
Given a pair $(|A|,\Delta)$,  \eqref{eq:2_reachable} defines the Bloch vector with the maximal distance on the Bloch sphere from the initial state that can still be reached, thus every state with an angle $\Theta$ smaller than this value can also be reached. Those states define the set of reachable angles. Note that for $\Delta=0$ we can reach all states on the Bloch sphere.
The time specified in Eq.~\eqref{eq:2_time} provides the time necessary to perform such a process if the desired Bloch vector is exactly at the boundary of the set of reachable states given by Eq.~\eqref{eq:2_reachable}. Finally, Eqs.~\eqref{eq:2_time} and \eqref{eq:2_phase} together provide conditions for the amplitude and phase of the control.
\subsection{State-to-state transfer for the star $N$-level system}
\label{sec:N-level}
For a generic $N$-level system, even with an acyclic graph, analytic solutions are rare. However, if we restrict the interaction furthermore such that only the transitions $\left| 0 \right\rangle \leftrightarrow \left| k \right\rangle\, (k \neq 0)$ are allowed and increasing the symmetry in the problem by setting all detunings to an equal value ($\Delta_{0k} \equiv \Delta \ \forall\, k \neq 0$), an analytic solution can be found. In the example presented in Fig.~\ref{fig:level_scheme_and_graph} this scenario occurs if from the complete graph the blue dashed edge between $E_1$ and $E_4$ is erased.\\
For a clearer display of the underlying dynamics we again assume the system initially to be in the ground state $\vec{c}\,(t=0) = (1,0,0, \dots , 0)$ of the drift Hamiltonian and define analogously to Sec. \ref{sec:2-level}
\begin{equation}
{A} \equiv \sqrt{\Delta^2 
+ \sum\limits_{f=1}^{N-1} |A_f|^2}.
\end{equation}
After some straightforward algebra, one finds
\begin{align}
c_0(t)& = e^{i\Delta /2 t}\left( \cos \frac{\tilde{A}}{2} t - i \frac{\Delta}{\tilde{A}} \sin  \frac{\tilde{A}}{2} t \right)\\
c_k(t) &= e^{-i\Delta /2 t} \frac{A_k}{\tilde{A}} \sin \frac{\tilde{A}}{2} t\quad (k = 1, 2, \dots , N-1).
\end{align}
To solve the state-to-state transfer problem we introduce the normalized goal state $
\vec{\xi} = (\xi_0, \xi_1 e^{i \beta_1}, \xi_2 e^{i \beta_2} \dots , \xi_{N-1} e^{i \beta_{N-1}})$, where $\xi_k$ and $\beta_k$ are real numbers and the global phase is chosen such that $\beta_0 = 0$. The procedure from here on and the structure of the resulting equations follow that of the two-level system: for a goal state to be reachable, the inequalities
\begin{equation}
\label{eq:N_reachable}
\frac{|A_k|}{\tilde{A}} \geq \xi_k
\end{equation}
have to hold for all $k \neq 0$. The total time $T$ is then fixed by
\begin{equation}
\label{eq:N_time}
T = \max_{k}\ \frac{2}{\tilde{A}} \arcsin \left(\xi_k \frac{|A_k|}{\tilde{A}} \right).
\end{equation}
Here, $T$ is again what is referred in the literature as the quantum speed limit. Furthermore the phases $\alpha_k$ of $A_k \equiv |A_k| e^{i \alpha_k}$ have to be chosen as
\begin{equation}
\label{eq:N_phase}
\alpha_k = \beta_k + \frac{1}{2} \Delta\, T.
\end{equation}
As before, the Eqs.~\eqref{eq:N_reachable}, \eqref{eq:N_time} and \eqref{eq:N_phase} determine if and how a goal state can be reached. While Eq.~\eqref{eq:N_reachable} refers to the reachability, Eqs.~\eqref{eq:N_time} and \eqref{eq:N_phase} state how the amplitudes and phases of the different $A_k$ have to be chosen to solve the problem provided that the goal state is at the boundary of the reachable set. Note that again for $\Delta=0$ we can reach all states.

\section{Numerical Results}
\label{sec:Numerics}
To bolster the analytic results obtained so far and show the frontier up to which the RWA is an excellent tool for QOC we simulate systems discusses in Sec.~\ref{sec:graphs}.
We create random non-degenerate energy spectra $\{E_0, E_1, \dots , E_{N-1}\}$, connected and acyclic Control Hamiltonians $\hat{H}_C$ and use the algorithm mentioned in Sec.~\ref{sec:graphs} to set up a time-independent Hamiltonian. As a figure of merit for state-to-state transfer we use the infidelity $I = 1 - \left| \left\langle \psi_G | \psi(T) \right\rangle\right|^2$. The parameters for the optimization are the amplitudes $\left|A_f\right|$, the phases $\alpha_f$ and the final time $T$. We fix the detuning $\Delta$ (which is assumed to be equal for all frequencies to study the effects of the detuning via changing one parameter instead of an increasing number while adding more freedom the task of optimization would be easier) and  optimize for many different random goal states. The optimization is performed via the direct search method Nelder-Mead~\cite{Nelder1965}.\\
To check if the results obtained in the RWA are valid in the complete description or if the RWA breaks down, we perform an exact time evolution with the results of the optimization and compared the results. Again, the infidelity is used as a figure of merit.\\
\begin{figure}[htp]
\begin{center}
\setlength{\unitlength}{.9cm}
\begin{picture}(7.5, 6.2)
\put(-0.6,0){\includegraphics[scale=0.64]{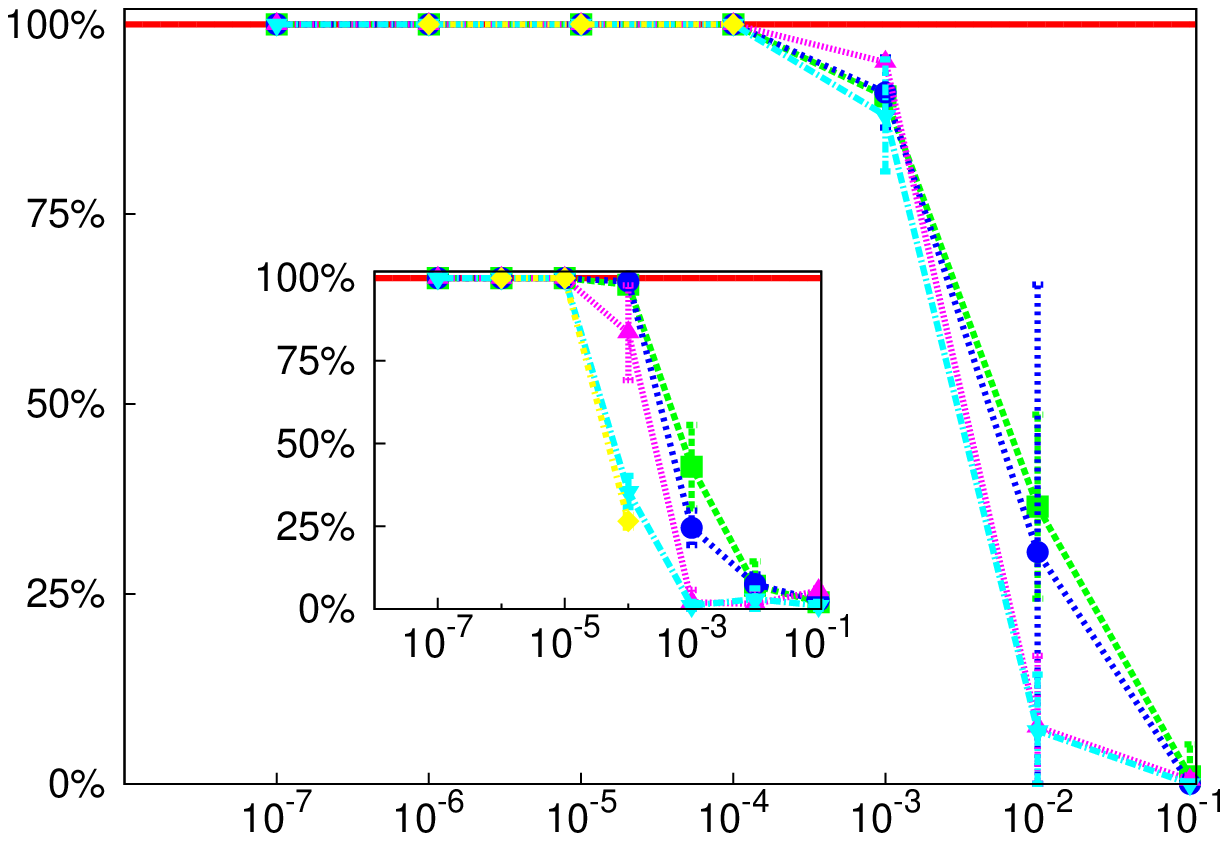}}
\put(4.1,-0.3){$\Delta$}
\put(-0.9,5.2){$\Lambda$}
\end{picture}
\caption{(Color online) Percentage $\Lambda$ of successfully optimized states as a function of detuning $\Delta$ for different dimension $N$: $2$ (green), $3$ (blue), $4$ (pink), $5$ (cyan) and $6$ (yellow). Inset: Percentage $\Lambda$ of positive double check as a function of detuning $\Delta$ for $N = 2, 3, 4, 5$ and $6$ with the same color scheme as the main figure.
}
\label{fig:percentage}
\end{center}
\end{figure}
\\
In Fig.~\ref{fig:percentage} the percentage of randomly generated states reached $\Lambda$ within an infidelity of less than $\epsilon = 10^{-3}$ is shown as function of the detuning. We see that it increases with decreasing detuning and that for every detuning smaller than $10^{-4}$ we are able to reach all states, that is the percentage of reached states is equal to $\Lambda = 100\%$. Moreover, for bigger dimensions the percentage of reached states decreases if the detuning is bigger than $10^{-4}$. The inset shows the result of the numeric double check, where we computed the time evolution of the exact system (without the RWA) and then calculated the infidelity of the exact time evolved state with respect to the goal state. We define a successful double check again as an infidelity below $\epsilon=10^{-3}$ and plotted it again as a function of the detuning $\Delta$ (see inset of Fig.~\ref{fig:percentage}). This allows us to see if the approximated model is still good enough for optimization. Remember that the double check can only be lower than or equal to the results from RWA, since a failed optimization within the RWA is very unlikely to produce a successful optimal pulse without the RWA. Accordingly a drop in the double check success rate indicates the break down of the RWA. The behavior is similar: For small detunings, that is less than $10^{-5}$ we have a $100\%$ quota of positive double checks; the bigger the detuning the smaller the number of positive double checks. Moreover we see that for bigger dimensions the success rate decreases, so for bigger dimensions the RWA needs smaller detunings to still be valid.
\begin{figure}[htp]
\setlength{\unitlength}{.9cm}
\begin{picture}(7.5, 6.5)
\put(-0.55,0.2){\includegraphics[scale=0.62]{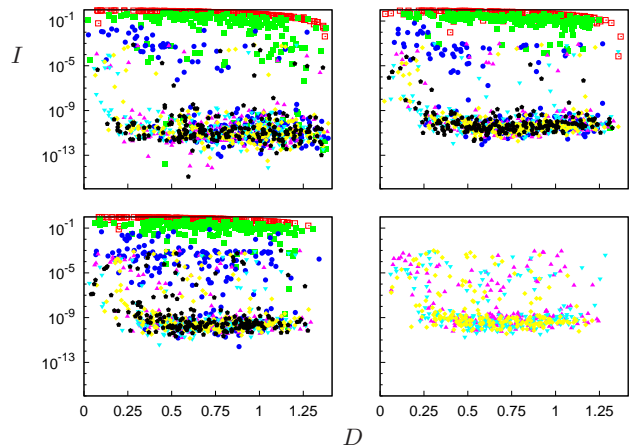}}
\put(-1,5.5){$I$}
\put(3.9,-0.15){$D$}
\end{picture}
\caption{(Color online) Reached infidelity $I$ as a function of the distance $D = \| \left| \psi_G \right\rangle - \left| \psi(0) \right\rangle\|$ for different detunings $\Delta$: $10^{-1}$~(red), $10^{-2}$~(green), $10^{-3}$~(blue), $10^{-4}$~(pink), $10^{-5}$~(cyan), $10^{-6}$~(yellow), $10^{-7}$~(black). Different pictures correspond to different dimensions $N$: $N~=~3$~(Top~left), $N~=~4$~(Bottom~left), $N~=~5$~(Top~right), $N~=~6$~(Bottom~right).
}
\label{fig:complete_data}
\end{figure}
In Fig.~\ref{fig:complete_data} we show all optimized states and different systems for different detunings (indicated by different colors, different plots correspond to different dimensions). On the x-axis we plot the distance between the initial state and the final state $D = \| \left| \psi_G \right\rangle - \left| \psi(0) \right\rangle\|$ in  Hilbert space, on the y-axis the infidelity $I$ is shown. We see that the reached infidelity for small detunings is far below the threshold of $\epsilon = 10^{-3}$ used for Fig.~\ref{fig:percentage}. Additionally we see the rapid drop in infidelity for smaller detunings and a bunching of data points around $10^{-10}$ or below for small detunings. In conclusion, we see that within the regime of a QOC task with very high fidelity in this important subset of possible problems can be performed without difficulties following the procedure laid out in this work.

\section{Example Application: Entangling Rydberg Atoms}
\label{sec:Example}
To demonstrate the presented approach in a real physical scenario we study the example of two trapped Rubidium atoms \cite{SaffmanReview,Gaetan,Urban,Isenhower,Zhang2010,Zhang2012}. For each atom we consider the qubit states $|0\rangle=|5S_{1/2},F=1\rangle$, $|1\rangle=|5S_{1/2},F=2\rangle$, and a Rydberg state $|r\rangle=|97D_{5/2}\rangle$ as in the experiments of Ref.~\cite{Zhang2012}. The atoms are trapped by spatially separated far-off-resonance optical traps and thus can be addressed individually by lasers. Both qubit states can be coupled to the Rydberg state via two-photon transitions, e.g. via $|5P_{1/2}\rangle$ and $|5P_{3/2}\rangle$. These intermediate states can be excluded from the model by appropriate laser detunings via adiabatic elimination~\cite{Brion2007}. The atoms interact only if both atoms are in the Rydberg state and the interaction is  $H_{int}=U|rr\rangle\langle rr|$ with $U=2\pi h\cdot20\,\mathrm{MHz}$, while the effective Rabi frequencies that couple the qubit states to the Rydberg state can be chosen to be a few MHz \cite{Zhang2012}.  The system was used to experimentally implement protocols for CNOT-gates \cite{Urban,Isenhower,Zhang2010,Zhang2012}. A similar setup has been used to propose a protocol for multi-particle entanglement \cite{Saffman2009} of the type $(|0...0\rangle+|1...1\rangle)/\sqrt{N}$.
Here we use the methods developed above to provide a solution how to transfer $|00\rangle$ to the Bell state $(|00\rangle + |11\rangle)/\sqrt{2}$ with a single (polychromatic) pulse, so in contrast to most schemes we do not make use of pulsed lasers used sequentially.

\begin{figure}[htp]
\begin{minipage}{0.25\textwidth}
\includegraphics[width=0.9\textwidth]{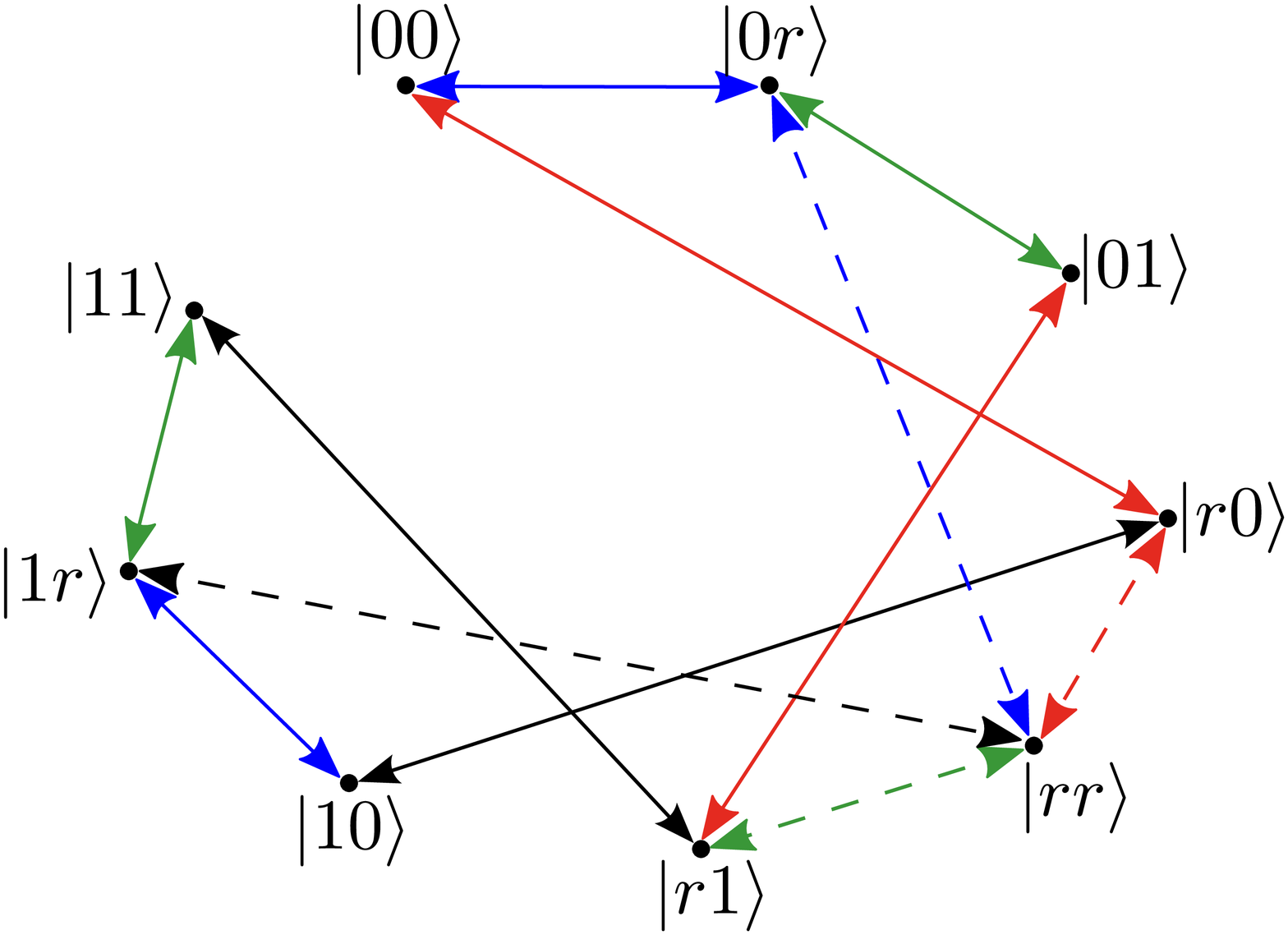}
\end{minipage}%
\begin{minipage}{0.25\textwidth}
\includegraphics[width=0.9\textwidth]{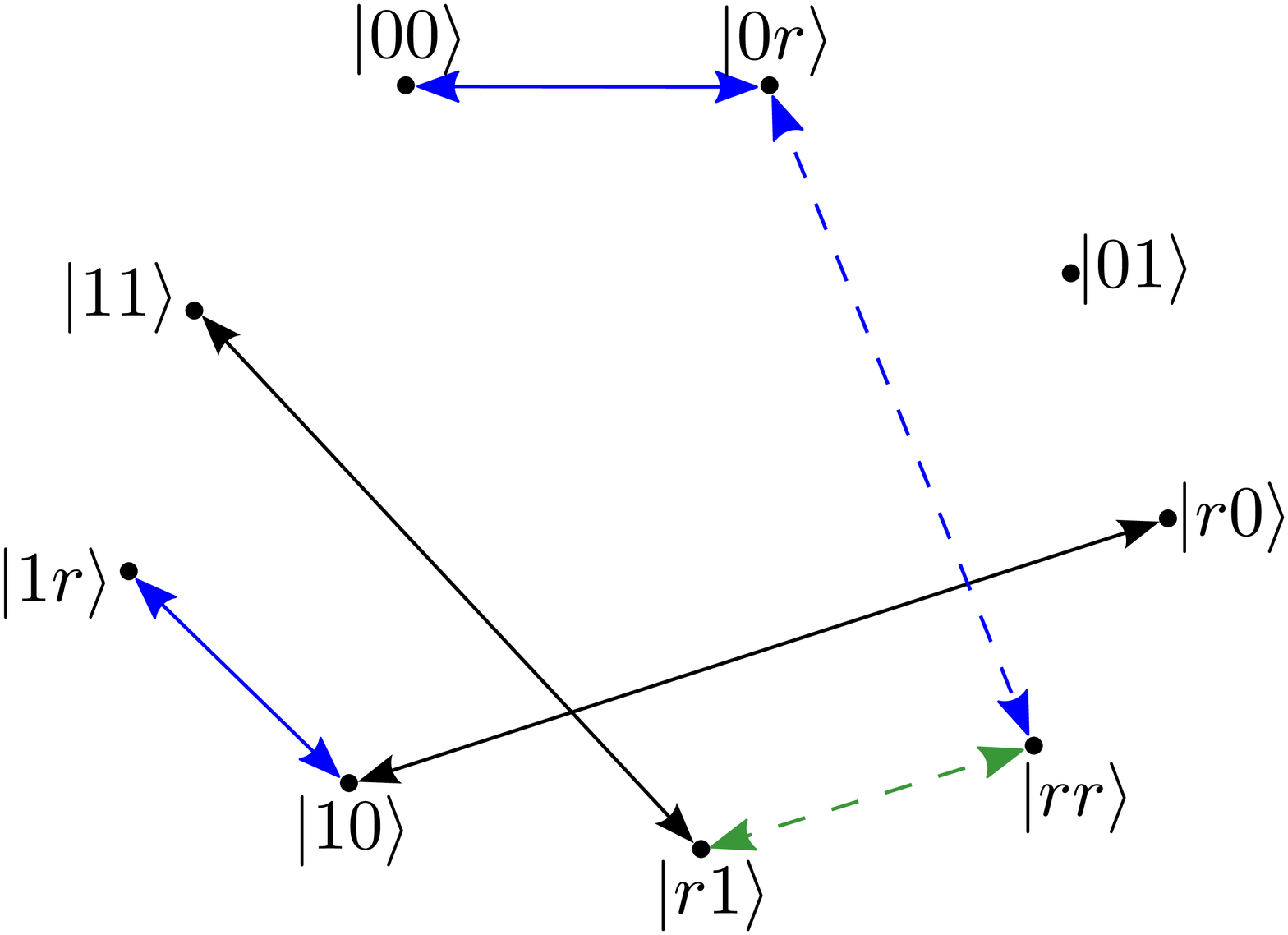}
\end{minipage}
\caption{(Color online) Left panel: The lasers couple the levels according to Eq.~\eqref{eq:rydberg}. Some lasers couple multiple transitions and the system is not fully controllable. Right panel: Using only four of the possible eight lasers, we are left with two controllable subsystems, one of them containing the initial state $|00\rangle$ as well as the state $|11\rangle$. On this subsystem we can apply the methods developed in this article to control the desired state-to-state transfer to the Bell state $(|00\rangle + |11\rangle)/\sqrt{2}$.
}
\label{fig:rydberg}
\end{figure}

If we perform the RWA and individually address the atoms, where we drive near-resonantly the transitions to the four states with a single Rydberg excitation as well as to the doubly excited Rydberg state (we double check the Rydberg blockade assumption numerically), the Hamiltonian reads
\begin{align}\label{eq:rydberg}
H&=\frac{\Omega_{1}}{2}(|00\rangle\langle 0r|+| 1r\rangle\langle  10|)
+\frac{\Omega_{2}}{2}(|00 \rangle\langle  r0|\nonumber\\
&+| 01\rangle\langle r1 |) +\frac{\Omega_{3}}{2}(|0r \rangle\langle  01| +|1r \rangle\langle 11 |)\nonumber\\
&+\frac{\Omega_{4}}{2}(|r0 \rangle\langle 10 | +| r1\rangle\langle 11 |)
+\frac{\Omega_{5}}{2}| 0r\rangle\langle  rr|\nonumber\\
&+\frac{\Omega_{6}}{2}| 1r\rangle\langle  rr|
+\frac{\Omega_{7}}{2}| r0\rangle\langle  rr|
+\frac{\Omega_{8}}{2}| r1\rangle\langle  rr|
+ h.c. \nonumber\\
&+\delta_{1} |0r\rangle\langle 0r|
+(\delta_{1} + \delta_{3}) |01\rangle\langle 01|
+\delta_{2} |r0\rangle\langle r0|\nonumber\\
&+(\delta_{1} + \delta_{5} + U) |rr\rangle\langle rr|
+(\delta_{1} + \delta_{2} + \delta_{3}) |r1\rangle\langle r1|\nonumber\\
&+(\delta_{2} + \delta_{4}) |10\rangle\langle 10|
+(\delta_{1} +\delta_{2} + \delta_{4}) |1r\rangle\langle 1r|\nonumber\\
&+(\delta_{1} +\delta_{2} + \delta_{3} + \delta_{4}) |11\rangle\langle 11|\,.\nonumber\\
\end{align}
This Hamiltonian leads to the graph that is shown in Fig.~\ref{fig:rydberg} (left). The single frequencies (same color and same line style in Fig.~\ref{fig:rydberg}) drive multiple transitions: e.g. the first laser ($\Omega_1$, blue solid line) drives the transition from $|00\rangle$ to $|0r\rangle$ as well as the transition from $|10\rangle$ to $|1r\rangle$. As a consequence the system is not fully controllable. However, if we switch only some of the lasers, namely $\Omega_1$, $\Omega_4$, $\Omega_5$, and $\Omega_8$, we get two controllable subgraphs (Fig.~\ref{fig:rydberg}, right). One of them contains our initial state $|00\rangle$ as well as our target state $(|00\rangle + |11\rangle)/\sqrt{2}$. If we solve the state-to-state transfer problem using experimental feasible values~\cite{Zhang2012} for the Rabi frequencies as well as the assumption of perfect blockade with the Hamiltonian~\eqref{eq:rydberg}, we can achieve perfect state transfer with $\Omega_1=2\pi h\cdot 1\,\mathrm{MHz}$, $\Omega_4=2\pi h\cdot 1\,\mathrm{MHz}$, $\Omega_5=2\pi h\cdot 3.2\,\mathrm{MHz}$, and $\Omega_8=2\pi h\cdot 1.3\,\mathrm{MHz}$, final time $T=314\,\mathrm{ns}$, $\delta_5=-U$ and all other Rabi frequencies vanishing. If we double check the assumption of perfect blockade and go beyond Eq.~\eqref{eq:rydberg}, the infidelity increases from $\epsilon=0$ to $\epsilon=0.002$. Note that this value can be improved if single addressing is possible for a smaller distance between the atoms, resulting in a higher blockade interaction. The operation error is thus just a technical limitations and not intrinsic to the method.

\section{Discussion and Conclusions}
\label{sec:DiscussionConclusion}
As we have seen in the previous sections QOC works very well within the RWA and clearly shows that the number of parameters for optimization has to scale linearly with the dimension $N$ of the Hilbert space describing the system~\cite{Lloyd2014,Moore2012}.
The intuitive argument for that is: Assume we add an additional frequency $\tilde{\omega}$ to those used to perform an already optimal protocol found previously. If $\tilde{\omega}$ is not in approximate resonance to any transition frequency it can be neglected and will not affect the description in the RWA. If it is in approximate resonance to any transition frequency, we just keep the new frequency $\tilde{\omega}$ and drop the old one which was on resonance with the same transition. 
In any case, there is no need to increase the number of frequencies beyond $N-1$.
This supports the findings of a recent work \cite{Lloyd2014} where, by means of an information theoretical analysis, it was shown that for an effective optimization the bandwidth of the control field (in this case given by the number of controls $F$) has to scale at least linearly with the dimension of the Hilbert space associated with the optimized system.

Another connection to a recently developed and highly efficient optimization algorithm, namely the Chopped Random Basis Algorithm (CRAB) \cite{CRAB}, can be made: In CRAB one expands the control pulse $u(t)$ into a truncated basis, often using trigonometric polynomials with great effectiveness. In a typically QOC problem, one has
\begin{equation}
\hat{H}_{\text{CRAB}} = \hat{H}_D + u(t) \hat{H}_{C},
\end{equation}
where
\begin{equation}
u(t) = \sum\limits_{n=1}^{N_c} A_n \sin \left( \omega_n t \right) + B_n \cos \left( \omega_n t \right).
\end{equation}
Analogously, one can rewrite the Hamiltonian \eqref{eq:system} in the form 
\begin{equation}
\hat{H} = \hat{H}_D + \sum\limits_{f=1}^{F} \big( \operatorname{Re} A_f\,  \cos \left( \omega_f t\right)\, +\, \operatorname{Im} A_f\, \sin\left( \omega_f t\right)\big) \ \hat{H}_C
\end{equation}
which shows a clear one to one relationship between the two methods. This once again backs the observations made with CRAB that the number of frequencies necessary for good optimization results is not exceedingly high, in particular it does not grow super-polynomially~\cite{Caneva2014,Lloyd2014}.

In conclusion, we investigated the performance of QOC within the generalized RWA applied to a $N$-dimensional quantum system. By introducing proper unitary transformations, we identified an important subset of QOC problems that can be described by a time-independent formalism, namely systems that can be described by acyclic, connected graphs.
We solved the state-to-state transfer problem for the paradigmatic two-level system, and the dynamics of the $N$-dimensional system whose graph is a star. We demonstrated numerically that a system representable by a connected, acyclic graph can be controlled to perform arbitrary state-to-state transfers and we showed that this approach allows to develop an optimal protocol to entangle Rydberg atoms with constant laser pulses, that is without the need of schemes for pulse shaping. Let us mention  that the subset of connected, acyclic graphs is of natural high interest for QOC since they 
represent the class of Hamiltonians that connect the complete $N$-dimensional system dynamics with the fewest possible controls, namely only $N-1$. Reducing the number of controls further either leaves the graph unconnected or introduces a near-degeneracy, both impeding optimal control.

To give an outlook, we stress that the class of QOC problems identified here, despite being quite general, does not include all scenarios where the presented approach can be applied successfully. In particular, even if the graph is not acyclic, there are cases in which the system can still be recast as a time-independent one. This is the case if the sum over all detunings along this cycle is zero: graphically this means that the phase accumulated is the same no matter which path of the cycle goes~\cite{Einwohner1976}. In Ref.~\cite{Whaley1984} a step further has been developed to encompass more of the complete system dynamics in a time-independent description by incorporating part of the counter-rotating terms, offering possibilities to enlarge the amount of applications even further.\\
\begin{acknowledgements}
We thank N. Rach and R. S. Said for useful discussions and comments.
This work was funded by SFB/TRR21 and the EU grants RYSQ and SIQS. This work was performed on the computational resource bwUniCluster funded by the Ministry of Science, Research and Arts and the Universities of the State of Baden-W\"urttemberg, Germany, within the framework program bwHPC.
\end{acknowledgements}

\end{document}